# EndoTOFPET-US
# a Novel Multimodal Tool for
# Endoscopy and Positron Emission Tomography

Erika Garutti

on behalf of the EndoTOFPET-US collaboration

*Abstract–* **The EndoTOFPET-US project aims to jointly exploit Time-Of-Flight Positron Emission Tomography (TOFPET) and ultrasound endoscopy with a multi-modal instrument for the development of new biomarkers for pancreas and prostate oncology. The paper outlines the functionality of the proposed instrument and the challenges for its realization. The high level of miniaturization and integration poses strong demands to the fields of scintillating crystallography, ultra-fast photon detection, highly integrated electronics and system integration. Solutions are presented to obtain a coincidence time resolution better than 200 ps and a spatial resolution of ~1 mm with an asymmetric TOFPET detector. A tracking system with better than 1 mm spatial resolution precision enables the online alignment of the system.**
**The detector design, the production and test status of the single detector components, and the integration plans are discussed.**

INTRODUCTION

PANCREATIC carcinoma is one of the most aggressive diseases and remains one of the most resistant cancers to current therapies. Its curative rate is very low due to unclear symptoms and the lack of specific biological markers (cite). Prostatic cancer on the other hand is the most diffuse cancer type affecting mankind, it can be cured with high probability of success if detected in an early stage and has the largest number of yearly cases treated.

The EndoTOFPET-US project [1] requires investigating specificity of the newly developed biomarkers with a spatial resolution of about 1 mm in the prostate or pancreas tissues. To achieve this the development of two novel detectors (depicted in Fig. 1) is mandatory: a PET head extension for a commercial ultrasound endoscope, and an outer PET plate to be placed outside the body in coincidence with the PET head. The two detectors need to be positioned with respect to each other to a precision of better than a millimetre for the duration of the examination. This task is performed by a combination of optic and electromagnetic tracking systems. Each of the two detectors consists of inorganic scintillators to convert the 511 keV photons from positron-electron annihilation to scintillation light. The crystals are coupled one-to-one to blue-sensitive photo-detectors. The signals of the photo-detectors are digitized either in the photo-detectors itself (internal probe), or via additional off-detector electronics (external plate). The system is controlled by an off-detector data acquisition and slow control system. The compact data acquisition system handles rates up to 40 MHz from the external plate and 200 kHz from the internal probe. The FPGAs on the front-end boards concentrate the event data from the external plate and transmit it to an external trigger. The data acquisition card then merges these data with the one from the internal probe. A pre-selection of coincidence candidates is performed directly in this trigger FPGA, while the more sophisticated event processing is handled by software.

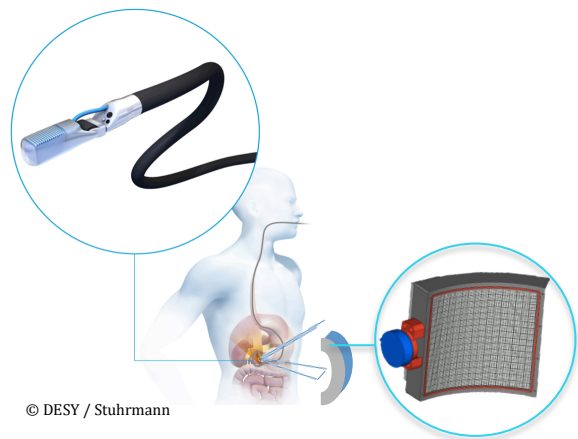

© DESY / Stuhrmann

Fig. 1. The EndoTOFPET-US detector system sketched in the application for pancreatic cancer diagnostics. Top left is the rendering of the ultrasound endoscope with PET extension, which needs to be positioned inside the stomach below the pancreas. Bottom right is the external plate detector positioned outside the body on the opposite position to the endoscope with respect to the pancreas.

Using commercial as well as in-house-developed silicon photomultipliers (SiPMs) the system aims to yield a coincidence time resolution (CTR) of < 200 ps FWHM needed

Manuscript received November 16, 2012. This work has been partly funded by the European Union 7[th] Framework Program (FP7/ 2007-2013) under Grant Agreement No. 256984 EndoTOFPET-US, and supported by a Marie Curie Early Initial Training Network Fellowship of the European Union 7[th] Framework Program (PITN-GA-2011-289355-PicoSEC-MCNet).

Erika Garutti is with the University of Hamburg, Luruper Chaussee 149, 22671 Hamburg, Germany (telephone: +49 040 8998-3779, e-mail: erika.garutti@desy.de).

for the background rejection from adjacent organs to the one under study. Modest energy resolution of order 20 % is sufficient to discriminate between photoelectric and Compton-scattered events. Minimum dead material between the crystals, in particular in the detector head is crucial to obtain acceptable sensitivity.

The extreme miniaturization of the detector head requires placing about 400 individual channels (crystals, photo-detector and readout electronics) in a volume of 23 mm diameter and 30-40 mm length. All these challenges are being addressed maintaining as a first priority that of meeting the requirements for a medical application on humans.

## I. MEDICAL REQUIREMENTS

The EndoTOFPET-US collaboration includes 6 European research institutes and Universities, 3 companies and three renowned University Clinical Centers with direct link to endoscopic facilities and extensive expertise in PET procedures. The clinical case for this novel multi-modal tool has long been established. Within this consortium the protocol for the final test of the tool in a medical environment was prepared considering anatomical-related limitations, maximum probe's size, shape and flexibility, organs uptake and background activity. In addition, specific operational constraints are identified like cleaning and disinfection procedures, international regulatory standards, definition of operating room workflow and hardware arrangement, and personnel exposure.

A project specific PET-CT database of > 850 anonym datasets have been collected for the system simulation. The collaboration is developing the software tools needed for the online image fusion of ultra-sound and PET imaging and the comparison to previously acquired PET-CT scans.

## II. SIMULATION OF AN ASYMMETRIC DETECTOR

The EndoTOFPET-US system is simulated using a custom-made adaptation of the GATE platform (Geant4 Application for Emission Tomography, [2]).

The official GATE package is not able to handle an asymmetric detector and has no capability of storing the step information from each single particle tracking through the detector a feature that is provided in GEANT4, but not propagated through GATE. The collaboration has fixed both problems and has developed a custom digitizer for the asymmetric detector system and an additional extension to trace the exact particle path in the detectors. The EndoTOFPET-US detector system geometry implemented in GATE is shown in Fig. 2.

For the simulation in the human environment the 4D phantom from NURBS XCAT* Cardiac Torso (WP Segars, John Hopkins Med Institution) is used. An additional problem is the necessity to place the internal probe inside the software phantom. GEANT4 does not offer the possibility to define overlapping volumes, therefore a smart solution has to be developed to position the internal probe for instance inside the stomach.

The first study performed with the adapted simulation package is the investigation of the system sensitivity. Fig. 3 presents the sensitivity of the system for a fixed internal probe geometry and various options for the crystal length and the dead space between crystals in the external plate. This study aims to guide in the selection of the best photo-detector package, so two existing package options are simulated. Using a SMD packaged monolithic array of $4 \times 4$ SiPMs (Hamamatsu model: MPPC S11828-3344M) the distance between crystals is kept at 200 μm. This improves the sensitivity of the detector by about 35 % compared to the design where single packaged SiPMs are used and the dead space between crystals is increased to more than 500 μm. The achievable sensitivity of the system is around 2000 cps/MBq.

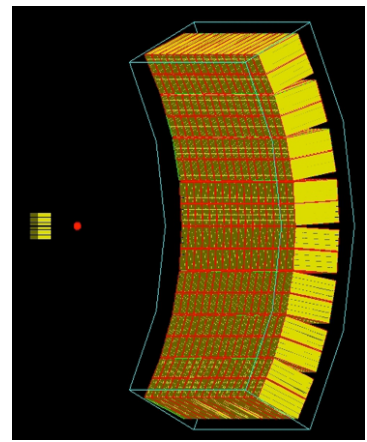

Fig. 2. The EndoTOFPET-US detector system geometry implemented in the GATE package [2]. Visible are the crystal matrices of the external plate (right) and of the internal probe (left). A point-like source in the vield-of-view of the two detectors indicates the typical distance of the organ under study from the two detectors.

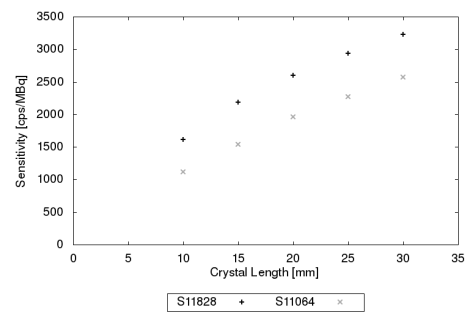

Fig. 3. The simulated sensitivity of the EndoTOFPET-US detector system using the GATE package [2]. Two geometries are tested for the external plate detector, corresponding to the achievable crystal separation when using one of the two possible photo-detectors considered for the project. Plus signs are for monolithic $4 \times 4$ MPPC (S11828) and crosses for an array of single MPPCs (S11064). A pointlike source of 0.5 mm radius is positioned 20 mm from the endoscope and 200 mm from the external plate. The simulation is performed for prostatic geometry of crystals (see text).

Further studies will be performed to determine the image resolution on point sources and extended sources in the phantom as a function of various detector geometries, detector performance and the possible implementation of depth of interaction readout.

## III. DETECTOR DESIGN

The EndoTOFPET-US detector system consists of two novel detectors, a PET head extension to a commercial ultra-sound endoscope, referred to as "internal probe", and its counterpart PET detector plate external to the human body.

### A. The internal PET probe

Depending on its use either as a trans-rectal (prostate) or gastrointestinal (pancreas) probe the detector head has two different sizes so as to adapt to the respective body orifices.

The detectors are made of thin ($0.71 \times 0.71$ mm$^2$) LYSO:Ce crystals grouped into a matrix of $9 \times 16$ crystals. One such matrix is used for the pancreas probe, two matrices are used for the prostate probe. Table I lists the details of the two PET probes. Each crystal is read out individually from one of its (small) sides. The crystal matrices are coupled via optical light concentrators to a digital silicon photo-multipliers (d-SiPMs) custom developed within the collaboration [10].

In contrast to what happens in an analog-SiPM, the sensitive pixels of a d-SiPM, or array of Single Pixel Avalanche Diodes (SPADs), are not summed to a common output, but read out by an individual counter on each pixel. Additionally, the pixels are column-wise connected to several TDCs to perform time-of-arrival (TOA) measurements on the first n detected photons. It was shown in [3] and [4] that this capability may guarantee to reach the Cramér-Rao [5], [6] lower bound for timing uncertainty no matter what threshold is chosen for the TOA measurement itself.

A good compromise between sufficiently large fill factor and a reasonable number of TOA measurements sets the number of TDCs per array to 48. The total fill factor of 57 % was achieved in the prototype production (0.35 μm CMOS test structure), considering the design trade offs between functionality and sensitivity.

Fig. 4 illustrates the technical design of the PET internal probe for the prostatic detector case. Visible are the two matrices of $9 \times 18$ crystals mounted on a common PCB hosting two SPAD arrays. The PCB is bent along one side and folded next to the crystals in the section of the probe. An FPGA is mounted on this interconnection piece, which serves for the communication of the SPAD arrays to the DAQ. Due to the power dissipation of the SPAD array and to the closed environment, cooling is needed to maintain the chips at a reasonably low temperature for their operation. Circulating room-temperature water through the cooling pipes in contact with the chip provides the necessary cooling. The detector is thermally isolated from the patient to prevent the contact with a too cold surface.

The position of the internal probe during the intervention is monitored using an electromagnetic tracking system capable of about 1 mm position resolution. The sensor (Aurora Mini 6DOF Sensor, $1.8 \times 9$ mm$^2$) is embedded in the internal probe support structure. An electromagnetic field generator (shown in Fig. 5) has to be positioned within the field of reach of the sensor (about 1 m$^3$) in the operating room. First calibration tests of the electromagnetic tracking systems are performed using a conventional optical tracking system as reference. In static conditions the precision of the tracking system exceeds the specified 1 mm accuracy. Measurements with a moving system are ongoing.

TABLE I: DETAILS OF THE TWO PET INTERNAL PROBES

|  |  | Prostate | Pancreas |
|---|---|---|---|
| Crystal Matrix | [mm$^2$] | 14 x 15 | 7 x 15 |
| Fiber length | [mm] | 10 | 10 |
| Fiber pitch x/y | [μm/μm] | 780/800 | 780/800 |
| # Fibers in x/y |  | 18x18 | 9x18 |
| SPAD array thick. | [mm] | 0.75 | 0.75 |
| PCB thickness | [mm] | 1 | 1 |
| # Readout layers |  | 1 | 1 |
| Total thickness | [mm] | 18 | 13 |
| Detector diameter | [mm] | 23 | 15 |
| Length of detector | [mm] | 22 | 22 |

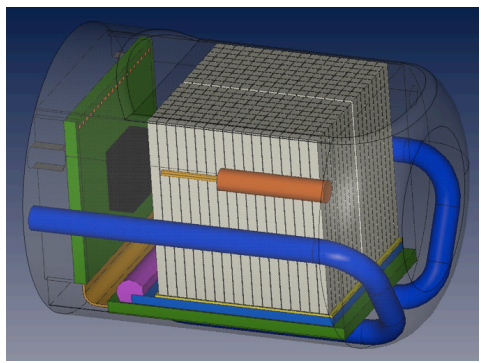

Fig. 4. Technical design of the PET internal probe for the prostatic detector case. Visible are the two matrices of 9x18 crystals mounted on a common PCB hosting the d-SiPM, or SPAD arrays. The readout is provided via the interconnection PCB hosting a control FPGA (black), which interfaces the SPADs to the DAQ. The two (blue) cooling lines maintain the PCB at room temperature. The electromagnetic tracking sensor (orange) is embedded in the support structure.

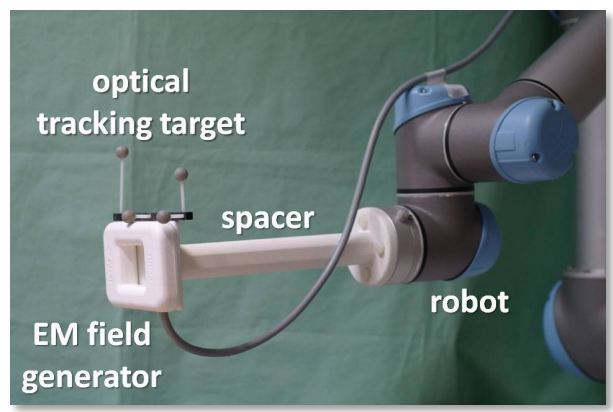

Fig. 5. Robotic arm under test for supporting the external plate detector. The electromagnetic field generator for the tracking sensor mounted on the internal probe is shown during a calibration measurement with a reference optical tracking system.

## B. The external PET plate

The external plate complements the internal probe and co-registers the back-to-back gammas from positron-electron annihilation. For an optimum coincident field of view, the plate is designed to have a size of 230 × 230 mm$^2$. It is made of matrices of 4x4 LYSO:Ce crystals, each with a volume of 3 x 3 × 15 mm$^3$. Each crystal matrix is coupled at the rear side to a 4x4 array analog MPPC[1]. Matrices are grouped into modules of 4 × 4 scintillators, for a total of 256 matrices or 4096 crystals.

For the readout of all these channels with the required time resolution of 200 ps in coincidence a dedicated ASIC chip has to be developed. The ASIC has to meet the additional requirements of very high channel density (~4000 ch / 20 × 20 cm$^2$), the possibility to tune the SiPM bias voltage for each individual channel by up to 0.5 V, and a low power consumption to minimize the cooling requirements.

Two solutions are currently being developed in parallel within the collaboration: the STiC-ASIC (STiC2v0 16 ch. prototype currently under test, [7] and the TOFPET-ASIC (64 ch. prototype submitted to IBM,[8]).

STiC is a mixed mode 16-channel ASIC chip in UMC 0.18 um CMOS technology aiming at SiPM readout with optimal timing resolution. It is designed for time of flight (ToF) measurements in HEP and medical imaging applications and dedicated in particular to the EndoTOFPET-US project. STiC provides either differential or single-ended input connection for SiPMs. The timing and charge information of the input signal are encrypted into two time stamps and processed later by an embedded TDC module with a timing resolution of better than 20ps. The digitized data are stored first on an on-chip SRAM block and then transferred out via a 160 MBit LVDS serial link using 8/10-bit encoding. The simulated single pixel timing jitter is less than 15 ps for Hamamatsu S10362-11-50 MPPCs. A special linearization method has been used to obtain a linear charge response in a very wide range. The total power consumption is less than 20 mW per channel.

The TOFPET-ASIC is a 25 mm² 64-channels chip designed in a standard CMOS 0.13 μm technology. One edge is free of pins, such that a rotated twin chip can be abutted to build a compact 128 channel module. A dual-threshold scheme allows to set a trigger level of 0.5 p.e., while a higher threshold is used both for SiPM dark count rejection and ToT measurements. Both timing and ToT measurements are performed by analogue TDCs based on time interpolation with a time binning of 50 ps. Event data is packed in frames and output by up to two LVDS data links, allowing for a maximum bandwidth of 640 Mbit/s. A power consumption between 5-10 mW per channel is expected to guarantee a signal-noise-ratio of at least 23.5 dB for the single photon, using a SiPM with 300 pF terminal capacitance.

## IV. SINGLE COMPONENTS CHARACTERIZATION

### A. Characterization of crystals

The prototypes of the crystal matrices for the internal probe and the external plate have been ordered from Proteus and from CPI respectively. They are shown in the top picture of Fig. 6. Both crystal types are LYSO:Ce and the reflector between crystals is ESR from 3M 80 μm thick. Both crystal matrices have been tested with a reference photo-detector.

Both matrices present a very good homogeneity of all crystals, and a good light yield of 10200 ph/MeV (11700 ph/MeV) for the external plate (internal probe) with the crystals in dry contact to the photodetector [9]. The bottom plot in Fig. 6 illustrates the good quality of the crytal matrix for the external plate. The energy resolution of 15% of the photo-peak is compatible with the required energy resolution of better than 20 %.

The time properties of the crystals have been tested using the analog SiPM envisaged for the external plate readout (MPPC 10931-050p, 3 × 3 mm$^2$, 50 × 50 μm$^2$ pixel size). The coincident time resolution (CTR) between two LYSO:Ce crystals of 3x3x15mm$^3$ and 0.71 × 0.71 × 10 mm$^3$ volume has been measured using the time over threshold method with the ultra-fast amplifier-discriminator chip NINO [11] in combination with a high precision TDC (HPTDC).

A CTR of 238.0 ± 4.3 ps FWHM is obtained. The same measurement repeated on the 16 crystals of the 4 × 4 matrix in coincidence with the same reference small crystal yields <CTR>= 250 ps with a spread of 20 ps [9].

---

[1] Two types of analog SiPM are currently under consideration: a monolithic 4 × 4 array with common cathode (S11828-3344M) and a single cathode array of 3 × 3 MPPCs with through via connections. Both products are produced by Hamamatsu Photonics.

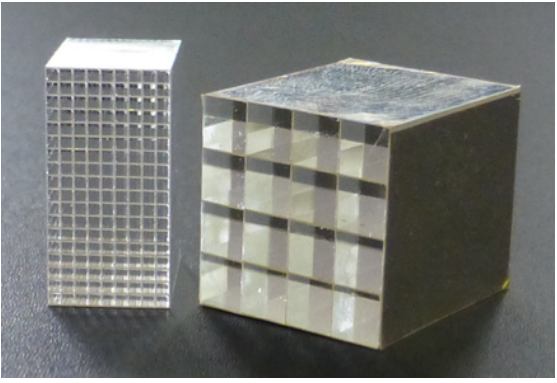

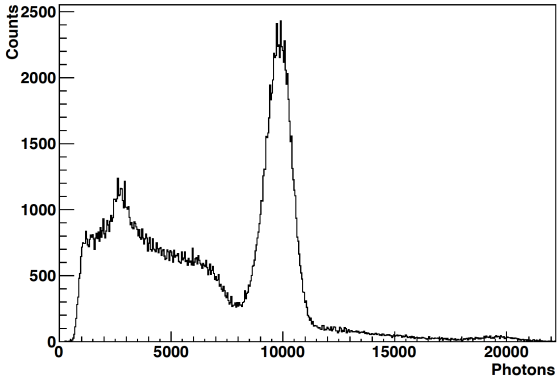

Fig. 6. The first two ptototypes of crystal matrices for the internal probe (from Proteus) and the external plate (from CPI) on top. Both crystal types are LYSO:Ce and the reflector between crystals is ESR from 3M. The bottom plot is the energy spectrum for the detection of 511 keV photons from $^{22}$Na obtained with the whole matrix of 16 crystals for the external plate [9].

### B. Characterization of digital SiPM

A prototype chip of SPAD arrays with 16 different geometries (see Fig. 7) has been produced and tested. Detailed results on the tests are reported in [10]. Out of the different structures on the chip one has been identified to be suitable for the application in the internal probe of the EndoTOFPET-US detector. The chosen structure is a matrix of 26 × 16 SPADs arranged in a single cluster. Each SPAD is equipped with a single bit counter for a total of 416 bits for the energy measurement. The fill factor of this structure is 57 % and the measured photo-detection efficiency for blue wavelength is 12.5 %. The SPAD array is operated typically 2-3 V above breakdown and the dark count rate (DCR) at room temperature is around 10 MHz, when all 416 pixels are enabled. The possibility to turn off individual (noisy) pixels to improve the SNR. The median DCR of 30-40 kHz was measured. The time measurement is performed using the 48 TDCs / SPAD array with time bins < 50 ps. The single pixel time resolution has been measured with blue laser light. Subtracting the laser jitter (35 ps) and the clock and PLL jitters of the readout test board a timing jitter of 115.0 ± 13.0 ps FWHM is obtained.

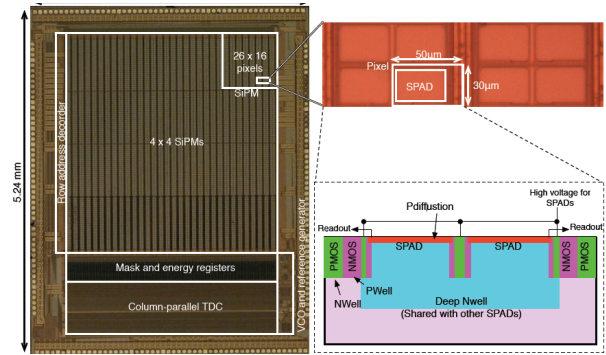

Fig. 7. Prototype SPAD array chip with 16 different geometry structures. The right insert zooms into the structure chosen for the EndoTOFPET-US internal probe detector. Single SPADs of 50 × 30 μm size are arranged in a matrix structure of 26 × 16 pixels [10]..

### C. Characterization of ASIC chip

A prototype version of the STiC-ASIC [7] has been produced in the 0.18 mm UMC CMOS technology and is currently under test. The chip provides energy measurements using the Time-over-Threshold (ToT) method and high time resolution measurements via an integrated TDC with 25 ps bins. The digital output is transmitted with a data rate of 160 Mbit/s.

From the first test results on the chip the digital configuration and serial data transmission are functional. The input bias DAC for the individual tuning of each SiPM is linear in the range of 0-0.7 V. The trigger jitter is found to be less than 30 ps for signals larger than 5 pC. The ToT output is linearly proportional to the input charge for signals larger than 3 pC, as demonstrated in Fig. 8. The linearity of the output is not a typical feature in ToT measurements, but it has been built in the ASIC design on purpose. If the input charge is large enough (> 3 pC) part of the input stage will be saturated and the charge is integrated on the detector capacitance. It is discharged through the input stage by a constant current, resulting in a linear pulsewidth - charge correlation.

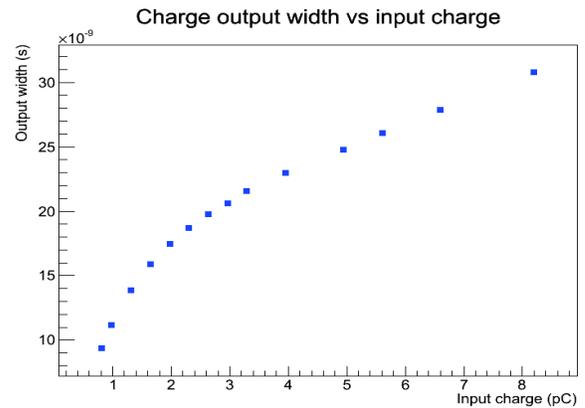

Fig. 8. Output width of the ToT measurement as a function of an input charge injected in the STiC2v02 prototype ASIC chip.

## V. INTEGRATION PLANS AND OUTLOOK

The selected Endoscope for the prostatic detector (Hitachi EUP-U533) has been selected and purchased. A mock-up of the internal probe is under construction. It will be connected to the Endoscope head without altering or modifying in any way the commercial Endoscope. Carbon fiber support for the PET head is being developed (see Fig. 9), which is mounted on a body of the Endoscope. The carbon fiber serves as envelop for all required cables for the SPAD arrays, the electromagnetic alignment sensor and the water-cooling connections. The first detector mock-up is in production and will be tested at the end of 2012. The end of 2013 expects the integration of the detector components described in this paper.

The commissioning and the clinical tests are foreseen for 2014.

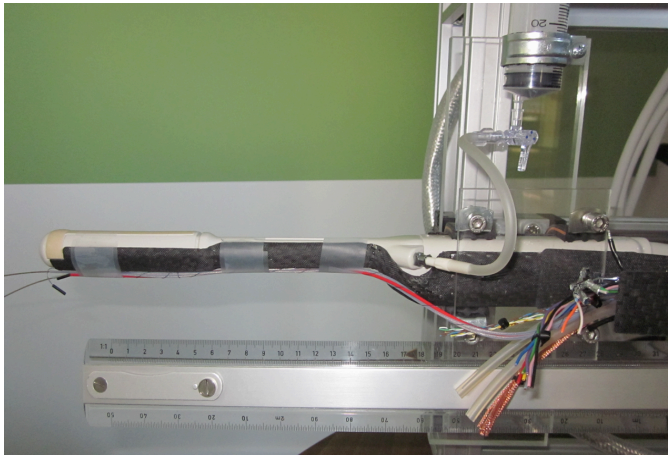

Fig. 9. The prostatic Endoscope Hitachi EUP-U533 covered by the carbon fiber support structure needed as fixation for the internal probe detector. Visible is also the electromagnetic sensor for the tracking of the internal probe.

ACKNOWLEDGMENT

The author wishes to express gratitude to the technical staff of the Collaboration for its invaluable support in building the test facilities and prototype devices used in the numerous measurements reported in this paper. She also wishes to thank her colleagues from all affiliated institutes for helping her prepare this paper.

This work has been partly funded by the European Union 7$^{th}$ Framework Program (FP7/ 2007-2013) under Grant Agreement No. 256984 EndoTOFPET-US, and supported by a Marie Curie Early Initial Training Network Fellowship of the European Union 7$^{th}$ Framework Program (PITN-GA-2011-289355-PicoSEC-MCNet)